\documentclass[12pt]{article}

\usepackage[dvips]{graphicx}

\begin{document}

\begin{center}
{\LARGE \bf Subspace Least Square Approach for Drift} \\
\vspace{5pt}
{\LARGE \bf Removal with Application to Herschel Data}

\vspace{20pt}
{\large Lorenzo Piazzo }

\vspace{20pt}
DIET Dept.  \\

Sapienza University of Rome  \\

Jan. 7, 2013 \\

\end{center}

\vspace{250pt}

\noindent
Lorenzo Piazzo \\
DIET Dept. - Sapienza University of Rome.  \\
V. Eudossiana 18, 00184 Rome, Italy.  \\
Tel.: +39 06 44585480 \\
Fax.: +39 06 4873300  \\
E-mail: lorenzo.piazzo@uniroma1.it  

\newpage

\tableofcontents

\newpage

\section{Introduction}

We consider the processing of the data produced by the two infrared imaging photometers onboard the ESA Herschel satellite \cite{herschel}, namely PACS \cite{pacs} and SPIRE \cite{spire}. These data are affected by several impairments and producing high quality images from the raw instrument output is a difficult task. In fact, in addition to the ubiquitous thermal noise, these data are also affected by offsets, saturations, pointing errors, glitches and drifts. As a result, the data are normally reduced by means of a pipeline composed of several steps, where each step takes care of a specific impairment and tries to remove it from the data and to produce a set of clean, updated data to be used as input for the next step. Tipically the last step receives data where all the impairments except the noise have been removed. The task of the last step is the production of a sky image (map) from the noisy data, a process called map making, which can be implemented using several well established methods, e.g. \cite{tegmark97}.

One of the impairments affecting the Herschel data is a time varying deviation from the baseline. This deviation, which is usually termed a drift, is typically slowly varying with respect to the signal and the noise but is normally much larger than the noise and often comparable to or larger than the signal component. Therefore it is mandatory to remove the drift before the map making step. This report is devoted to presenting a drift removal method suitable for use with Herschel data. Specifically, we assume that the drift can be represented as a polynomial and develop a Subspace Least Square (SLS) approach to its estimation, using concepts and techniques from linear algebra. Subspace analysis is a well known technique and it was used in the past do perform drift estimation, e.g. \cite{fmri}, but, to the best of our knowledge, the specific approach proposed here is novel. Moreover subpace based drift estimation seems to have not been applied to Herschel data before. The method presented here is employed in the Unimap map maker\footnote{More precisely, Unimap exploits a smart, iterative implementation of the method presented here. This implementation will be described in a future work.} \cite{unimap}.

While our main application is the reduction of Herschel data, we note that the drift can be found also in other types of data, for example in fMRI data \cite{fmri}. Moreover the approach presented in this report is not limited to a polynomial drift, but can handle also other drift models: it suffices that the drift can be modeled as a linear combination of given waveforms and that the data is a redundant, linear observation. Therefore the dedrifting method presented here could be exploited for other types of data too. For this reason the presentation is divided in three parts. The first part, section \ref{sec_pre}, is devoted to a brief summary of basic facts from linear algebra that are used in the report. In the second part, section \ref{sec_sls}, we introduce the SLS method, without any reference to a specific data type. In the third part, section \ref{sec_herschel}, we specialise the method to the case of the polynomial drift affecting the Herschel data.

\section{Preliminaries}
\label{sec_pre}

\subsection{Notation}

We use uppercase letters to denote matrices, lowercase letters to denote vectors and scalars.  A superscript $T$ denotes matrix or vector transposition. We say that a matrix with $N$ rows and $M$ columns is a $N \times M$ matrix. We use $I$ to denote the identity matrix. We use $E \{ x \}$ to denote the expected value of $x$. 

\subsection{Basic facts from linear algebra}
\label{sec_basic}

Let us summarise some basic facts which will be useful and can be found in any textbook covering linear algebra, e.g. \cite{meyer}. We will work in the vector space $\Re^N$ of the $N \times 1$ real column vectors but the results are easily extended to the complex case. In this space, the inner product of two vectors $x$ and $y$ is the scalar product $x^Ty$, the two vectors are said orthogonal if $x^Ty = 0$ and a set of vectors $v_i$ for $i=1,...,n$ is said to be an independent set if $\sum_i a_i v_i = 0$ only when the scalars $a_i$ are all zero.

Given a vector space $V$, a subspace $\Sigma \subset V$ is any subset of $V$ that is closed under vector sum and multipication by a scalar, i.e. such that if $x \in \Sigma$ and $y \in \Sigma$ then $(ax + by) \in \Sigma$. Note that a subspace is a vector space. 

Given a space $V$, a set of its vectors $B = \{ v_1, ..., v_n \}$ is said a basis if it is an independent set and if any vector $x$ of $V$ can be obtained as a linear combination of the vectors of $B$, i.e. it can be written as $x = \sum_i a_i v_i$ for some scalars $a_i$. The dimension of a subspace is the number of vectors in one of its basis.

The intersection of two subspaces $\Sigma$ and $\Delta$, denoted by $\Sigma \cap \Delta$, is a subspace and is constituted by all the vectors belonging to both $\Sigma$ and $\Delta$. If $\Sigma \cap \Delta = 0$ (the null vector) the subspaces are said disjoint.

The sum of two subspaces $\Sigma$ and $\Delta$, denoted by $\Gamma = \Sigma + \Delta$, is a subspace and is constituted by all the vectors that can be written as $x = s + d$ where $s \in \Sigma$ and $d \in \Delta$. If $\Sigma$ and $\Delta$ are disjoint they are said to be complementary in $\Gamma$ and $\Gamma$ is said the direct sum of $\Sigma$ and $\Delta$, denoted by $\Gamma = \Sigma \oplus \Delta$. In this case a vector in $\Gamma$ can be written as $x = s + d$ in a unique way (i.e. if also $x = w + h$ where $w \in \Sigma$ and $h \in \Delta$ then $w =s$ and $h =d$).

Consider $\Gamma = \Sigma \oplus \Delta$ and one of its vector $x = s + d$. The vector $s \in \Sigma$ is said the projection of $x$ in $\Sigma$ along $\Delta$ and the vector $d \in \Delta$ is said the projection of $x$ in $\Delta$ along $\Sigma$. It exists a matrix $\Pi$ such that $s = \Pi x$ and $d = (I-\Pi)x$. The matrix $\Pi$ is termed the projector into $\Sigma$ along $\Delta$ and the matrix $(I-\Pi)$ the projector into $\Delta$ along $\Sigma$. All the projectors are idempotent, i.e. $\Pi^2 = \Pi$. Conversely all the idempontent matrices are projectors into some subspace.

Given a subspace $\Sigma$ of a space $\Gamma$ the set of all the vectors in $\Gamma$ that are orthogonal to all the vectors of $\Sigma$ is a subspace termed the orthogonal complement of $\Sigma$ in $\Gamma$ and is denoted by  $\Sigma^\perp$. Note that $\Gamma = \Sigma \oplus \Sigma^\perp$, i.e. $\Gamma$ can be regarded as the direct sum of any of its subspace and the corresponding orthogonal complement. Therefore a vector of $\Gamma$ can be uniquely written as $x = s + s^\perp$ where $s \in \Sigma$ and $s^\perp \in \Sigma^\perp$. In this case $s$ is the orthogonal projection of $x$ into $\Sigma$ and $s^\perp$ is the orthogonal projection of $x$ into $\Sigma^\perp$.

Consider a set of $n$ vectors $M = \{ v_1, ..., v_n \}$. The set of all the vectors that can be obtained as a linear combination of the $v_i$, i.e. all vectors than can be written as $x = \sum_i a_i v_i$ for some scalars $a_i$, is a subspace and is said the span of $M$.

Given an $n \times m$ matrix $X$ the span of its columns is called the range of $X$ and is a subspace of $\Re^n$ denoted as $\Sigma_X$. Note that a vector $y$ in the range can be written as $y = X a$ where $a$ is a $m \times 1$ vector. Another important subspace is the null of the matrix, which is a subset of $\Re^m$, constituted by all the vectors $a \in \Re^m$ such that $Xa = 0$.

Consider an $n \times m$ matrix $X$ such that its columns are an independent set. Consider the range $\Sigma_X$ and the orthogonal complement $\Sigma_X^\perp$. Then any vector $x \in \Re^n$ can be decomposed as $x = s + s^\perp$ where  $s \in \Sigma_X$ and $s^\perp \in \Sigma_X^\perp$. The orthogonal projection of $x$ into $\Sigma$, namely the vector $s$, is given by $s = \Pi_X x$ where $\Pi_X$ is a $n \times n$ matrix given by $\Pi_X = X ( X^T X ) ^{-1} X^T$ which is termed the orthogonal projector into $\Sigma_X$. Similarly, the orthogonal projection of $x$ into $\Sigma^\perp$, namely the vector $s^\perp$, is given by $s^\perp = \Pi^\perp_X x$ where $\Pi^\perp_X = I - \Pi_x$. The null of $\Pi_X$ is $\Sigma_X^\perp$ and the null of $\Pi_X^\perp$ is $\Sigma_X$. Any orthogonal projector is idempotent and symmetrical, i.e. $\Pi_X = \Pi_X^T$.

\section{Subspace least square drift estimation}
\label{sec_sls}

\subsection{Data model and problem statement}
\label{sec_prob}

Consider a signal vector $m \in \Re^{M}$ which is observed with a linear instrument to produce a vector of $N >> M$ observed data $s = Pm$ where $P$ is an $N \times M$ matrix representing the linear instrument. Assume that the observations are affected by two additive distrubances. The first is a random $N \times 1$ vector $n$ representing thermal noise. The second is a $N \times 1$ vector $y$ representing a drift in the data. In the following we assume that the drift can be expressed as $y = X a$ where $X$ is a $N \times K$ known matrix, with $K<<N$ and $a$ is an unknown $K \times 1$  vector. In words the condition requires that the drift is a linear combination of the columns of $X$. Then the noisy observation, denoted by $d$, is a vector written as
\begin{equation}
d = Pm + Xa + n = s + y + n.
\label{noisy_data}
\end{equation}

We consider the problem of producing an estimate $y^*$ of the drift vector $y$. Then the drift estimate can be subtracted from the data vector to produce an updated data vector $\tilde{d} = d - y^*$ where, ideally, the drift has been removed. Such an updated data vector can next be used to estimate the signal $m$. In the model of (\ref{noisy_data}) we assume that the matrices $P$ and $X$ are known, that the vectors $m$ and $a$ are unknown deterministic vectors and that $n$ is a random process. We also assume that $P$ and $X$ are full rank, because this simplifies the presentation and is true for our main application, i.e. Herschel data, but it would not be difficult to remove such hypothesis. Also, we make the development assuming real numbers, but the extension to complex numbers is not difficult.

\subsection{Noiseless case analysis}
\label{sec_noiseless}

As a preliminary step we consider how to produce a drift estimate $y^*$ when the noise is absent, i.e. we set $n=0$ in (\ref{noisy_data}). This case is simpler to analyse but allows to introduce all the tools that we need in the solution of the noisy case. 

One problem in the drift estimation is the presence of the signal component $s$ in the data. In fact the presence of the signal will bias the drift estimate. In order to solve this problem we consider the span of the matrix $P$, which is a subspace $\Sigma_P$ that will be termed the signal subspace, and the corresponding projector, which is $\Pi_P = P ( P^T P ) ^{-1} P^T$. We also consider the orthogonal complement $\Sigma^\perp_P$, which will be termed the nosignal subspace, and the corresponding projector $\Pi^\perp_P = (I - \Pi_P)$. Next we note that $s = Pm$ lies in the signal subspace so that $\Pi^\perp_P s = 0$. Then we can get rid of the signal by projecting $d$ into the nosignal subspace. In fact the projection is $z = \Pi^\perp_P d$ and, since $d = s + y$,  we obtain
\[
z = \Pi^\perp_P y 
\]
showing that the signal has been removed. 

Now we note that the drift vector belongs to the subspace spanned by the columns of $X$, which will be denoted by $\Sigma_X$ and termed the drift subspace. Then an estimate of the drift can be obtained by selecting a vector $y^* \in \Sigma_X$ that, after being projected into the nosignal subspace, is equal to $z$. Since any vector $v$ in $\Sigma_X$ can be written as $v = Xa$, after the projection the vector will be in the form $\Pi^\perp_P X a = W a$, where we introduced the $N \times K$ matrix $W = \Pi^\perp_P X $. Then we can look for $y^*$ by solving the following equation in the variable vector $a$
\begin{equation}
W a = z.
\label{eq1}
\end{equation}
Once we have a solution $a^*$, the drift estimate is $y^* = X a^*$. We now proceed to better discuss the solution of the last equation.

Since we assumed that the noise is absent, the last equation has at least one solution, because $z$ is indeed the projection of a drift vector into the nosignal subspace. However if $\Sigma_X \cap \Sigma_P \neq 0$, there will be more than one solution. In fact, suppose that $a^*$ is a solution and $y^* = X a^*$ is the corresponding drift estimate, such that $\Pi^\perp_P y^* = z$. Now consider any vector $v \in \Sigma_X \cap \Sigma_P$. Since $v \in \Sigma_X$, it can be written as $v = X b$ for some vector $b$. Moreover, since $v \in \Sigma_P$, we have $\Pi^\perp_P v = 0$. Then it is not difficult to show that $a^* + b$ is also a solution. In fact
\[
W (a^* + b) = W a^* + \Pi^\perp_P X b = z + \Pi^\perp_P v = z.
\]
The last reasoning shows that the general solution of the system is in the form $a^* + b$ where $a^*$ is a particular solution and $b$ is such that $v = X b$ is in the signal subspace. We now proceed to discuss a method to find a particular solution of (\ref{eq1}).

As we have seen equation (\ref{eq1}) has at least one solution. This means that $z$ lies in the subspace spanned by the columns of the matrix $W$, which will be denoted by $\Sigma_W$. We have also seen that, when $\Sigma_X \cap \Sigma_P \neq 0$, the equation has infinitely many solutions. This means that $z$ can be expressed in more than one way as a linear combination of the columns of $W$ implying that the columns of $W$ are a dependent set. We now present a procedure to remove columns from the matrix $W$ to construct a matrix $\hat{W}$ such that the columns of $\hat{W}$ still span $\Sigma_W$ and are an independent set. In other words the columns of $\hat{W}$ are a basis for $\Sigma_W$. The procedure also produces a partition of $\Sigma_X$ into the direct sum of two subspaces, denoted by $\Sigma_{\hat{X}}$ and $\Sigma_{\bar{X}}$  that will be useful later. 

The procedure is started by initialising $k=0$, $\hat{X}^{(0)} = X$ and $\bar{X}^{(0)}$ to a void matrix. Next the following steps are iterated.

\noindent
1. Check if $\Sigma_{\hat{X}^{(k)}} \cap \Sigma_P = 0$. If true set $\hat{X} = \hat{X}^{(k)}$, $\bar{X} = \bar{X}^{(k)}$ and stop.

\noindent
2. Identify a non zero vector $v \in \Sigma_{\hat{X}^{(k)}} \cap \Sigma_P$.

\noindent
3. Since $v \in \Sigma_{\hat{X}^{(k)}}$ we have $v = \sum_i b_i x_i$ where the $x_i$ are the columns of $\hat{X}^{(k)}$ and the $b_i$ are appropriate coeffiecients not all equal to zero. 

\noindent
4. Select a non zero coefficient. Suppose it is the $j$-th one.

\noindent
5. Set $\hat{X}^{(k+1)}$ equal to $\hat{X}^{(k)}$ with the $j$-th column removed.

\noindent
6. Set $\bar{X}^{(k+1)}$ equal to $\bar{X}^{(k)}$ with the vector $v$ appended as last column.

\noindent
7. Let $k = k+1$. Go to step $1$.

As a preliminary comment, note that the procedure can be repeated no more than $K$ times, because if it is repeated $K$ times the matrix $\hat{X}^{(K)}$ will be a void matrix, the test in step 1 will be positive and the procedure will stop. In practice the procedure will be repeated $\alpha$ times where $\alpha$ is the dimension of the subspace $\Sigma_X \cap \Sigma_P$.

The output of the procedure is constituted by the $N \times (K - \alpha)$ matrix $\hat{X}$ and by the $N \times \alpha$ matrix $\bar{X}$. The span of $\hat{X}$ is a subspace $\Sigma_{\hat{X}}$ that will be termed the drift-nosignal subspace. The span of $\bar{X}$ is a subspace $\Sigma_{\bar{X}}$ that will be termed the drift-signal subspace. Furthermore we can consider the $N \times (K - \alpha)$ matrix $\hat{W} = \Pi^\perp_P \hat{X}$. These matrices and spaces are important in the development, because of the features summarised in the following lemmas which are proved in the appendix.

{\bf Lemma 1.} The columns of the matrix $\hat{W} = \Pi^\perp_P \hat{X}$ are an independent set.

{\bf Lemma 2.} The columns of the matrix $W = \Pi^\perp_P X$ are in the span of the columns of $\hat{W}$, i.e. they are all obtainable as linear combinations of the columns of $\hat{W}$.

{\bf Lemma 3.} We have $\Sigma_X = \Sigma_{\hat{X}} \oplus \Sigma_{\bar{X}}$, $\Sigma_{\hat{X}} \cap \Sigma_P = 0$ and $\Sigma_{\bar{X}} = \Sigma_X \cap \Sigma_P$.

We now discuss how to find a particular solution of system (\ref{eq1}). To this end consider the following, reduced system in the variable $(K-\alpha) \times 1$ vector $\hat{a}$
\begin{equation}
\hat{W} \hat{a} = z
\label{eq2}
\end{equation}
and note that, since the columns of $\hat{W}$ are a subset of those of $W$, a solution $\hat{a}^*$ of the reduced system immediately brings a solution of the full system of (\ref{eq1}). The full solution $a^*$ can be obtained by completing $\hat{a}^*$ with some zero coefficients, placed in correspondence of the columns of $W$ that are missing in $\hat{W}$. Then, in order to find a particular solution of the full system, we need to solve the reduced system.

The reduced system surely has a solution because $z$ is in the span of $\hat{W}$ due to Lemma 2. Moreover, it is not difficult to verify that the reduced system only has one solution, because $\Sigma_{\hat{X}} \cap \Sigma_P = 0$. Since the reduced system has only one solution, the solution is also the unique solution of the associated system of normal equations \cite{meyer} 
\begin{equation}
\hat{W}^T \hat{W} \hat{a} = \hat{W} ^T z
\label{eq3}
\end{equation}
and since, from Lemma 1, we know that the columns of $\hat{W}$ are an independent set, implying that the matrix $( \hat{W}^T \hat{W}  )$ is non singular, the solution is 
\[
\hat{a}^* = ( \hat{W}^T \hat{W}  )^{-1} \hat{W}^T z.
\]

Let us now develop an expression for the drift estimate. Since we have an expression for a solution of the reduced system, we can construct a solution, $a^*$, of the full system by completing with zero coefficients and compute the drift estimate as $y^* = X a^*$. However it is immediate to verify that $y^*$ can be obtained directly from the solution of the reduced system as
\[
y^* = \hat{X} \hat{a}^*.
\]
Moreover, by replacing $\hat{a}^*$, $z = \Pi_P^\perp d$ and $\hat{W} = \Pi_P^\perp \hat{X}$ and by using the fact that $\Pi_P^\perp$ is symmetric and idempotent, we can eventually write the drift estimate as 
\begin{equation}
y^* = \hat{X} (\hat{X}^T \Pi_P^\perp  \hat{X})^{-1} \hat{X}^T \Pi_P^\perp d = H d
\label{drift_est}
\end{equation}
where we introduced the matrix $H = \hat{X} (\hat{X}^T \Pi_P^\perp  \hat{X})^{-1} \hat{X}^T \Pi_P^\perp$. The latter expression gives a drift estimate that depends only on the data vector $d$ and on the matrices $P$ and $X$.

\subsection{Subspace Least Square drift estimation}

Let us now consider the drift estimation when the noise is present. As before, we can get rid of the signal by projecting the data vector in the nosignal subspace. Therefore, as before, we compute $z = \Pi_P^\perp d$. 

Again a meaningful drift estimate would be obtained by selecting a vector in $\Sigma_X$ that, after being projected into the nosignal subspace, is equal to $z$. To this end we should look for a vector $a$ such that
\[
Wa = z.
\]
However, since $z$ is now a noisy vector, in general an exact solution to the preceeding overdetermined system does not exist. An obvious way out is to find a least square solution, i.e. to look for a vector $a$ that minimizes $|Wa-z|^2$. However the least square solution is not unique, because, again, it is not dificult to show that given a solution $a^*$ and a vector $v = Xb  \in \Sigma_X \cap \Sigma_P$ then also $(a^*+b)$ is a solution. Like already done in the noiseless case, since the span of $W$ and $\hat{W}$ is the same, this problem can be solved by replacing the original minimization with the minimization of $|\hat{W} \hat{a}-z|^2$, with the advantage that the latter minimization only has one solution, which can be obtained as the solution of the associated system of normal equations. The normal equations are identical to those of the noiseless case given by (\ref{eq3}) so we find that the optimum $\hat{a}$ is still
\[
\hat{a}^* = ( \hat{W}^T \hat{W}  )^{-1} \hat{W}^T z
\]
and obtain a corresponding drift estimate, $y^* = \hat{X} \hat{a}^*$, which is identical to (\ref{drift_est}), namely
\[
y^* = Hd.
\]
Then the latter expression gives the drift estimate also in the noisy case and will be referred as the Subspace Least Square (SLS) drift estimate. 

Finally, by subtracting $y^*$ from $d$ an updated data vector is obtained as
\begin{equation}
\tilde{d} = d - H d = (I - H) d.
\label{data_upd}
\end{equation}
In the next section we better characterize the updated data vector and the drift estimate. 

\subsection{Analysis of the SLS estimate}

In order to better characterize the SLS estimate we need to study the matrix $H$. To this end we first note that, as is easy to check, $H^2 = H$ so that $H$ is idempotent. Therefore $H$ is a projector. We now proceed to make a convenient partition of $\Re^N$. Specifically let us consider the subspace which is obtained by the sum of $\Sigma_P$ and $\Sigma_{\hat{X}}$, denoted by $\Sigma_D = \Sigma_P + \Sigma_{\hat{X}}$. Since, from Lemma 3, $\Sigma_P \cap\Sigma_{\hat{X}} = 0$ this is in fact a direct sum, i.e. $\Sigma_D = \Sigma_P \oplus \Sigma_{\hat{X}}$. Next let us introduce the orthogonal complement of $\Sigma_D$, denoted by $\Sigma^\perp_D$. Since $\Re^N = \Sigma_D \oplus \Sigma^\perp_D$ we get the following decompositon for $\Re^N$
\[
\Re^N = \Sigma_P \oplus \Sigma_{\hat{X}} \oplus \Sigma^\perp_D
\]
so that every vector $v$ in $\Re^N$ can be uniquely written as
\begin{equation}
v = v_P + v_{\hat{X}} + v_{D^\perp}
\label{deco}
\end{equation}
where $v_P \in \Sigma_P$, $v_{\hat{X}} \in \Sigma_{\hat{X}}$ and $v_{D^\perp} \in \Sigma_{D^\perp}$.

We now study the multiplication of the matrix $H$ with a vector taken from one of the three subspaces just introduced. We firstly consider a vector $v \in \Sigma_{\hat{X} }$. This vector can be written as $v = \hat{X}b$ for some vector $b$ so that
\begin{equation}
H v = \hat{X} (\hat{X}^T \Pi_P^\perp  \hat{X})^{-1} \hat{X}^T \Pi_P^\perp \hat{X} b = \hat{X} b = v  \hspace{2cm} v \in \Sigma_{\hat{X}}.
\label{v0}
\end{equation}
Now consider a vector $v \in \Sigma_P$, which can be written $v = P b$ for some vector $b$. We have
\begin{equation}
H v = \hat{X} (\hat{X}^T \Pi_P^\perp  \hat{X})^{-1} \hat{X}^T \Pi_P^\perp P b = 0  \hspace{2cm} v \in \Sigma_{P}
\label{v1}
\end{equation}
where we used the fact that $\Pi_P^\perp P = 0$. Finally consider a vector $v \in \Sigma_D^\perp$. This vector is orthogonal to all the vectors of $\Sigma_P$, therefore it is in $\Sigma_P^\perp$ so that $\Pi_P^\perp v = v$. This vector is also orthogonal to all the vectors of $\Sigma_{\hat{X}}$ so that $\hat{X}^T v = 0$. Then
\begin{equation}
H v = \hat{X} (\hat{X}^T \Pi_P^\perp  \hat{X})^{-1} \hat{X}^T \Pi_P^\perp v = \hat{X} (\hat{X}^T \Pi_P^\perp  \hat{X})^{-1} \hat{X}^T v = 0  \hspace{2cm} v \in \Sigma_D^\perp.
\label{v2}
\end{equation}
The last three equations show that $H$ is the projector into $\Sigma_{\hat{X}}$ along $\Sigma_P \oplus \Sigma_D^\perp$. Conversely, $(I-H)$ is the projector into  $\Sigma_P \oplus \Sigma_D^\perp$ along  $\Sigma_{\hat{X}}$.

We are now ready to discuss the drift estimate and updated data vector. The drift estimate is $y^* = H d$ where $d$ is given by (\ref{noisy_data}). Let us now express the noise, drift and signal components using the writing of (\ref{deco}). The noise can be written as $n = n_P + n_{\hat{X}} + n_{D^\perp}$. The signal lies in the signal subspace, so that $s = s_P$. Concerning the the drift $y$, note that, based on Lemma 3, it can be written as $y = y_{\hat{X}} + y_{\bar{X}}$ where $y_{\hat{X}} \in \Sigma_{\hat{X}}$ and $y_{\bar{X}} \in \Sigma_{\bar{X}}$. However, from the same Lemma, we also have $y_{\bar{X}} \in \Sigma_P$. Therefore the drift can be written as $y = y_{\hat{X}} + y_{P}$. Now, using these writings and equations (\ref{v0}, \ref{v1}, \ref{v2}), it is easy to check that 
\[
y^* = y_{\hat{X}} + n_{\hat{X}}. 
\]
The last equation shows that the SLS estimate is the drift component falling into the drift-nosignal subspace while the component falling into $\Sigma_P$ is not detected. The estimate is affected by a random error given by $n_{\hat{X}}$. Note that $E \{ n_{\hat{X}} \}= E \{ H n \} = 0$ if the noise is zero mean.  

Let us now discuss the updated data vector which, from (\ref{data_upd}), can be written as
\[
\tilde{d} = \tilde{s} + \tilde{y} + \tilde{n} 
\]
where $\tilde{s} = (I-H)s$ is the updated signal,  $\tilde{y} = (I-H)y$ is the updated drift and $\tilde{n} = (I-H)n$ is the updated noise. Since $s \in \Sigma_P$ it will not be modified by $(I-H)$ so that 
\[
\tilde{s} = s
\]
telling us that the signal part is unmodifed in the updated data vector. Now consider the drift. Using $y = y_{\hat{X}} + y_P$ we get 
\[
\tilde{y} = y_{P}
\]
telling us that the drift component falling into the signal subspace will leak into the updated data vector. Then the best case is when the intersection of the signal and drift subspaces only contains the zero vector, because in this case $\tilde{y} = 0$ and the drift is entirelly removed. More generally the drift estimation will be good and the approach useful as long as most of the drift energy falls outside the signal subspace. 

Finally, using $n = n_P + n_{\hat{X}} + n_{D^\perp}$, the updated noise is 
\[
\tilde{n} = n_P + n_{D^\perp}.
\]
Note that the updated data vector will be passed to a noise removal algorithm and the latter equation shows that the noise component falling in the signal subspace is not changed in the updated data vector. This means that the performance of the noise removal algorithm will not be affected by the drift removal. In fact any good noise removal method should not be affected by the noise falling outside the signal subspace. Also, for the updated noise we have
\[
E \{ \tilde{n} \}= E \{ (I-H) n \} =  (I-H) E \{ n \} = 0,
\]
showing that if the original noise is zero mean the same is true for the updated noise, and 
\[
R_{\tilde{n}} =  E \{ \tilde{n} \tilde{n}^T \}= (I-H) E \{  n n^T \} (I-H)^T  =  (I-H) R_n (I-H)^T ,
\]
showing the correlation matrix of the updated noise can be computed from the correlation matrix of the original noise.

\subsection{Direct and Iterative SLS}
\label{sec_ite}

Let us briefly discuss some implementation issues. The updated data vector is obtained by subtracting the drift estimate from the original data vector. Then we only discuss the computation of the drift estimate.

The drift estimate can be computed directly, using (\ref{drift_est}). This requires performing a mutliplication with the matrix $H = \hat{X} (\hat{X}^T \Pi_P^\perp  \hat{X})^{-1} \hat{X}^T \Pi_P^\perp$. The multiplication can be carried out in successive steps: we firstly multiply the vector by $\Pi_P^\perp$; then we multiply the result by $\hat{X}^T$; and so on. The most difficult step is the multiplication by  $(\hat{X}^T \Pi_P^\perp  \hat{X})^{-1}$ since this requires producing the inverse of $(\hat{X}^T\Pi_P^\perp  \hat{X})$, which may not be feasible if the dimension of the matrix is large. 

An alternative way of producing the drift estimate is using an iterative method. For example, we can solve the normal equations of (\ref{eq3}) using the Parallel Conjugate Garadient (PCG) \cite{numrec} method. The PCG method requires no matrix inversion but only to perfom the multiplication of a vector by the matrix $\hat{W}^T$ or $\hat{W}$, which is a much simpler task.

Having concluded the general presentation of the SLS approach, in the next sections we study how this approach can be applied to the specific case of the Herschel data.

\section{Application to Herschel data}
\label{sec_herschel}

\subsection{Herschel data model}
\label{sec_data}

In order to use the SLS approach for Herschel data, we need to cast the Herschel data into the model of equation (\ref{noisy_data}) and derive the matrices $P$ and $X$. To this end we firstly discuss the data acquisition process.

The PACS and SPIRE instruments onboard the Herschel Satellite are imaging photometers made by arrays of bolometers, measuring the power emitted in several infrared bands. The number of bolometers in the array will be denoted by $N_b$ and varies depending on the instrument and on the observation band.

The arrays observe a field of view in the sky, covering an area of about $6$ square arcmin. However a typical Herschel observation covers a larger sky area, which may be some square degrees wide. To observe the area, the Herschel telescope is moved along one or more sets of parallel scan lines. During the scan each bolometer is sampled to produce a sequence of $N_r$ readouts which is termed a timeline. The set of all the $N_b$ timelines, toghether with the corresponding pointing information, constitutes the observation raw output. By stacking the timelines, the oservation output can be represented compactly with a $N \times 1$ vector $d$ where $N = N_b \cdot N_r$.

We assume that all impairments except the drift and the noise have been removed in prior processing steps. Then the data vector can be written as $d = s + y + n$ where $s$ is the signal, $y$ is the drift and $n$ is the noise. In the rest of this section we deepen the noise and signal models, while the drift component will be discussed in the next section. 

The noise component is typically modeled as a zero mean, stationary Gaussian process with a power spectrum given by the sum of two terms. The first term is white noise with flat spectrum $N_w(f) = N_0$, the second term is a correlated noise with spectrum $N_c(f) = ( f_k / f)^{\alpha} N_0 $ where $f_k$ is called the knee frequency. This term is also referred as the $1/f$-noise and dominates the spectrum at low frequencies, causing long lasting departures from the zero level.

A model for the signal component is obtained by assuming that the observed sky is a pixelised image, i.e. assuming that the sky is partitioned into a set of $M$ non overlapping squares (pixels) and that the flux is constant in each pixel. Then, by stacking all the pixels, the observed sky can be represented as a $M \times 1$ vector $m$, which is termed a map. Since each readout is accompanied by a pointing information we can assign each readout to a pixel, i.e. each element of vector $d$ to an element of vector $m$. This correspondance can be cast into a $N \times M$ matrix $P = \{p_{k,i} \}$, termed the pointing matrix, such that $p_{k,i} = 1$ if the $k$-th readout falls into the $i$-th map pixel and  $p_{k,i} = 0$ otherwise. In this way the signal component can be expressed as $s = Pm$, which fits into the model of (\ref{noisy_data}). Note that $P$ is a sparse matrix the rows of which are all zero except for one element which is one. 

As a comment note that in the model just introduced we must consider some limits to the pixel size. Specifically the noise and drift removal algorithms need redundancy in order to do a good job. In practice this requires that each pixel is observed several times so that $N >> M$.  Therefore the pixel size shall be large enough to guarantee that condition, a fact which puts a lower limit to the resolution of the final image. 

As a further comment note that each timeline is affected by an unknown offset because the instruments' readouts are not absolute but relative. Tyipically, this offset is roughly removed in the first processing step, by forcing each timeline mean or median to zero. The fine offset compensation is carried out by the image formation algorithm (the map maker) which follows the drift removal. However the absolute offset cannot be estimated and only the relative timelines' offsets can be corrected. As a result the final image itself is affected by an unknown offset that has to be estimated using independent calibration data.

\subsection{Herschel drift model}

The drift component of a single timeline is normally well approximated by a polynomial of low degree, say less that five. In order to develop a model for the drift let us initially assume that there is only one timeline, with $N_r$ samples, which is modeled as a polynomial of degree $N_a$. In this case the elements of the drift vector can be written as $y_i = \sum_{k=0}^{N_a-1} x_i^k a_k$ for $i=1,...,N_r$, where $a_k$ for $k=0,...,N_a$ are the polynomial coefficents and $x_i$ are real numbers. Since the sampling of the bolometer is done at regular times, the $x_i$ shall be equispaced numbers, of the form $x_i = i \delta + \alpha$. Now note that 
\[
y_i = \sum_{k=0}^{N_a} x_i^k a_k = \sum_{k=0}^{N_a} (i \delta + \alpha)^k a_k 
\]
and by developping the powers in the last expression and then grouping the powers of $i$ we can write
\[
y_i = \sum_{k=0}^{N_a} i^k b_k  
\]
where the $b_k$ are appropriate coefficients. The last expression shows that the values $y_i$ when $x_i = i \delta + \alpha$ can also be obtained as a polynomial with coefficents $b_k$ when $x_i = i$. Then, without loss of generality, we can assume\footnote{Note that, in the practical implementation of SLS, we shall select $\alpha$ and $\delta$ in order to improve numerical stability.} $\delta = 1$ and $\alpha = 0$, i.e. that $x_i = i$. We now introduce the $N_r \times (N_a+1)$ matrix $\tilde{X}$ given by
\[
\tilde{X} = \left(
\begin{array}{lllll}  
1 & x_1 & x_1^2 & ... & x_1^{N_a}  \\
1 & x_2 & x_2^2 & ... & x_2^{N_a}  \\
1 & x_3 & x_3^2 & ... & x_3^{N_a}  \\
...  & ...    & ... & ...    &  ...          \\
1 & x_{N_r} & x_{N_r}^2 & ... & x_{N_r}^{N_a}  
\end{array} \right),
\]
the $(N_a+1) \times 1$ vector $a$
\[
a = \left(
\begin{array}{l}  
a_0   \\
a_1  \\
...        \\
a_{N_a}  
\end{array} \right)
\]
and note that the drift can be written as 
\[
y = \tilde{X}a
\]
which fits the data model of (\ref{noisy_data}). We also note that $\tilde{X}$ is a Vandermonde matrix so that its columns are linearly independent when $N_r \geq (N_a+1)$. 

The model just introduced is easily expanded to the case when there are $N_b$ timelines of $N_r$ samples, each affected by independent polynomial drifts, with polynomial order $N_a$. In this case the vector $d$ is obtained by stacking the timelines and similarly can be done with the drift component $y$. Then, upon denoting the $h$-th timeline drift as a vector $y^{(h)}$, using the development of the previous paragraph, this vector can be expressed as $y^{(h)} = \tilde{X} a^{(h)}$ where $a^{(h)}$ is a column vector with the drift coefficients of the $h$-th timeline. Furthermore, by introducing a block diagonal matrix of dimension $N \times K$, with $N = N_b \cdot N_r$ and $K = N_b \cdot (N_a+1)$, given by 
\[
X = \left(
\begin{array}{lllll}  
\tilde{X} & 0 & 0 & ... & 0  \\
0 & \tilde{X} & 0 & ... & 0  \\
0 & 0 & \tilde{X} & ... & 0  \\
...  & ...    & ... & ...    &  ...          \\
0 & 0 & 0 & ... & \tilde{X}  
\end{array} \right)
\]
the drift vector can be written as 
\[
y = X a
\]
where $a$ is a $K \times 1$ vector obtained by stacking the $a^{(h)}$ vectors. 

To proceed we note that the map making algorithm employed to produce the final image may be capable of removing the drift too, at least to a certain extent. In fact the map maker is essentially a noise removal algorithm and the drift can be seen as a low frequency noise. In this case there is no need to separately estimate and remove the drift. However we also note that the drift observed in Herschel data often has common components. Typically one can identify a common drift, affecting all the timelines, in addition each timeline's drift. Furthermore, when the array is partitioned into subarrays, like in PACS, a subarray drift component is present. Such common components are not well removed by the map maker, because the mapper normally assumes that the noise processes of the timelines are uncorrelated, which is not true in the presence of the common drift. In this case we can run the derifting algorithm to remove the correlated drift components only and leave to the mapper the burden of the single drift removal. Theferore we now proceed to generalise the drift model just introduced to handle the case of common drift components.

In order to develop a model for the common drift components, let us assume that the timelines are divided into $N_g$ groups and that each group is affected by a polynomial drift with order $N_a$. It is not difficult to construct a matrix $X$ and a vector $a$ suitable to model this situation. Without loss of generality we can assume that all the timelines of a group are stacked successively in the data vector. Now consider the $h$-th group and denote by $g_h$ the number of timelines in the group and by $a^{(h)}$ the vector of the coefficients of the drift. All the timelines in the group are affetcted by this drift therefore we introduce the vector $y^{(h)}$ obtained by stacking $g_h$ copies of the drift, one for each timeline in the group. Then we stack $g_h$ copies of the  matrix $\tilde{X}$ to produce a matrix  $X^{(h)}$ given by
\begin{equation}
X^{(h)} = \left(
\begin{array}{l}  
\tilde{X}    \\
\tilde{X}   \\
...			\\
\tilde{X} \\
\end{array} \right)
\label{Xh}
\end{equation}
so that we can write $y^{(h)} = X^{(h)} a^{(h)}$ to model the drift of the group. The complete model is obtained by stacking all the $y^{(h)}$ vectors into a single $N \times 1$ vector $y$ which is the drift, with $N = N_b \cdot N_r$, by stacking all the $a^{(h)}$ vectors into a single $K \times 1$ vector $a$, with $K = N_g \cdot(N_a+1)$, by introducing the $N \times K$ matrix 
\begin{equation}
X = \left(
\begin{array}{lllll}  

X^{(1)} & 0 & 0 & ... & 0  \\
0 & X^{(2)} & 0 & ... & 0  \\
0 & 0 & X^{(3)} & ... & 0  \\
...  & ...    & ... & ...    &  ...          \\
0 & 0 & 0 & ... & X^{(N_g)}  
\end{array} \right)
\label{x2}
\end{equation}
and by noting that $y = X a$, which fits into the model of equation (\ref{noisy_data}). Note that when $N_g = N_b$ we are back to the case of a single drift per timeline, therefore the last model is the most general and the only one we need to study.

\subsection{Projecting in the signal subspace}
\label{sec_proj}

Having developed a suitable model for the Herschel data, in the next sections we consider several practical and theoretical issues concerning the implementation of the SLS drift estimate for Herschel data. As a starting point we discuss the projection of the data vector into the nosignal subspace, i.e. the computation of $z = \Pi^\perp_P d$, since this is the first step in the method. In practice, $z$ can be obtained as $z = d - v$ where $v$ is the projection of $d$ into the signal subspace, $v = \Pi_P d$. Then it is sufficient to discuss the projection into the signal subspace.

The projector into the signal subspace is $\Pi_P = P ( P^T P )^{-1} P^T$ which, by introducing the matrix $R = ( P^T P )^{-1} P^T$, can be written as $\Pi_P = P R$. The projection of $d$ into the signal subspace can thus be written as $v = P R d$. Now consider the vector $r = Rd$. It is a $M \times 1$ vector which can be regarded as an estimate of the map (in fact it is the LS estimate) and is termed the naive or simple projection or rebinned map. To better understand how the naive map is obtained note that it is not difficult to verify that the vector $P^Td$ is a map where the $i$-th element is equal to the sum of all the readouts falling into the $i$-th pixel and that $( P^T P )$ is a $M \times M$ diagonal matrix where the $i$-th diagonal element is equal to the number of readouts falling into the $i$-th pixel. Then it is seen that the $i$-th element of $r$ is equal to the average of all the readouts falling into the $i$-th pixel. Now we can express the projection as $v = P r$ and note that this operation, termed the back-projection of the $r$ map, amounts at producing the data vector that would be obtained if the sky was given by the vector $r$. In summary, the projection of a data vector $d$ into the signal subspace amounts at firstly rebinning $d$ into a naive map and next at back-projecting a data vector from the naive map. 

Both the projection and backprojection operations are simple to perform and can be efficently implemented without really producing and storing the matrix $P$. Therefore the projection into the signal and nosignal subspaces poses no implementation problems. Also note that if one of the diagonal elements of $(P^TP)$ is zero then the matrix is singular, but this means that there is an unobserved pixel in the map and this can be solved by simply removing that pixel from the problem. Therefore in a well posed problem the $(P^TP)$ matrix is non singular. 

\subsection{SLS for Herschel data}
\label{sec_ana}

Let us discuss the drift removal capability of the SLS approach for the specific case of Herschel data. To this end recall that the SLS approach will not remove the drift components falling in $\Sigma_{P}$ while the rest is perfectly removed. Therefore it is useful to study the intersection of the signal and the drift subspaces, spanned by the columns of the $P$ and $X$ matrices respectively. This discussion is also useful in order to implement the procedure described in section \ref{sec_noiseless}, where we remove columns from the matrix $X$ to produce the matrix $\hat{X}$.

As a first point note that a constant $N \times 1$ vector $c = (1, 1, ..., 1)^T$ lies in both subspaces. In fact when $X$ is given by (\ref{x2}), the constant vector can be obtained as $c = X a$ by properly selecting the coefficent vector $a$, so that $c$ lies in the drift subspace. Now consider the projection of the $c$ vector into the signal subspace, namely the vector $t = \Pi_P c$. As we have seen $\Pi_P = P R$, i.e. it can be seen as a naive mapping followed by a back-projection. Then, by rebinning $c$ we obtain $r = R c$ which is easily shown to be a constant map, $r = (1,1,...,1)^T$. Next, it is easy to verify that by back-projecting the constant naive map we obtain a vector $t = Pr$ which is the constant vector, i.e. $t=c$. Then $c$ is projected onto itself by $\Pi_P$ and is therefore also in the signal subspace.

Since the constant vector is in both subspaces, it is in the intersection of the two, together with all its scaled versions. Whether the intersection contains other vectors or not depends on the particular $P$, i.e. on the scan strategy and the pixel size, and cannot be discussed in general. However, since all the vectors in $\Sigma_X$, are obtained by stacking polynomials, as soon as the redundancy is moderately high, so that several redouts fall into each pixel, it is highly unlikely that one of these vectors also belongs to the $\Sigma_P$ subspace, since it should pass unchanged through $\Pi_P$, i.e. through a naive mapping and a backprojection operation. Therefore for all practical cases we can safely assume that the intersection contains the constant vector and no other vectors.

Since $\Sigma_X \cap \Sigma_P$ only contains constant vectors and since all other drift components are correctly estimated by the SLS, neglecting the noise the drift estimate can written as $y^* = y + c$ where $c$ is an unknown constant vector and $y$ is the actual drift. The latter equation says that, for practical Herschel data, the drift estimate is affected by an unknown offset but is otherwise exact. On the other hand the presence of an offset is no real problem, because we have seen in section \ref{sec_data} that the production of absolutely calibrated maps is impossible without the use of independent calibration data. Therefore if the drift estimation is affected by an offset, it makes no difference. In this sense we can say the the SLS approach entirelly removes the drift and is ideally suited for Herschel data.

Finally, let us discuss how to produce the matrix $\hat{X}$. To this end note that $\Sigma_X \cap \Sigma_P$ only contains the constant vector and that the constant vector is obtained by combining with the same coefficient all and only the columns of $X$ where the first column of $\tilde{X}$ is replicated. Then, according to the procedure described in section \ref{sec_noiseless}, to produce $\hat{X}$ we can remove anyone of these columns. The simplest choice is to remove the first column of $X$, which is one of such columns.

\subsection{Drift removal with Direct and Iterative SLS}

The direct estimation of the drift can be obtained using (\ref{drift_est}) and requires to compute $Hd$. Since $H = \hat{X} (\hat{X}^T \Pi_P^\perp  \hat{X})^{-1} \hat{X}^T \Pi_P^\perp$ the multiplication can be carried out in successive steps, without really computing and storing the $H$ matrix, which would be unfeasible for Herschel data. The first step is to mutliply $d$ by $\Pi_P^\perp$, which can be efficently implemented with naive mapping and back-projection, as we have seen in section \ref{sec_proj}. Then we multiply the result, which is a $N \times 1$ vector, by $\hat{X}^T$ and note that the matrix $\hat{X}$ is obtained from $X$ of (\ref{x2}) by removing one column, as we have seen in section \ref{sec_ana}. Then it is a $N \times (K-1)$ sparse, block matrix, constructed from the $\tilde{X}$ matrix. Then $\hat{X}$ does not really need to be produced and stored: the multplication can be efficently carried out storing only $\tilde{X}$, with a computational complexity similar to naive mapping. The result of the last step is a $(K-1) \times 1$ vector which needs to be multiplied with the $(K-1) \times (K-1)$ matrix $(\hat{X}^T \Pi_P^\perp  \hat{X})^{-1}$. This is the most delicate step and will be discussed in the next paragraph. It produces a $(K-1) \times 1$ vector containing the drift coefficients. The last step is the multplication of this vector by $\hat{X}$. This amounts at the synthesis of the drift of each timeline which again can be efficently implemented using the $\tilde{X}$ matrix and has a complexity similar to naive mapping.

Let us now discuss the multiplication by $(\hat{X}^T \Pi_P^\perp  \hat{X})^{-1}$. We found no clever ways to implement this step and need to use the brute force approach. That is we produce the matrix $(\hat{X}^T \Pi_P^\perp  \hat{X})$ and compute and store its inverse. The production of the matrix can be efficently realised and is normally feasible. The result is a $(K-1) \times (K-1)$ matrix where $K = N_g \cdot (N_a+1)$ which we need to invert. Whether this is feasible depends on $K$. As we mentioned the drift is well modeled by a low order polynomial, say less than five, then, in practice, $K$ depends on the number of groups $N_g$. For example suppose we want estimate a drift for each timeline, so that $N_g = N_b$, in a PACS blue observation using a polynomial order of $3$. Since $N_b = 2048$ for PACS blue we have $K = 2048 \cdot 4 = 8192$ and the matrix is too big to be inverted. On the other hand if we only want to remove the subarray drift, since PACS has $8$ subarrays we only need to compute the drift for $N_g=8$ groups and $K = 8 \cdot 4 = 32$ which yields a matrix that can easily be inverted. In practice the use of direct SLS is limited to the removal of the common drifts. The iterative approach based on the PCG needs to be used when $K$ is too high for the direct approach.

\subsection{Projecting in the drift subspace}

To conclude, let us discuss the implementation of the projection into the drift subspace, $\Pi_X = X (X^TX)^{-1}X^T$. Note that this projection is not needed in the context of the SLS as presented thus far. However it will be useful for a future, planned development, therefore we discuss it here.

 The projection $\Pi_X d$ can be implemented in steps. The first step is multiplying $d$ by $X^T$ and we have already seen that this step can be implemented efficently. Similarly the last step, i.e. the multiplication by $X$, is simple and has been already discussed. The only step worth discussion is the multiplication by $(X^TX)^{-1}$ which is discussed in the next paragraph.

As a preliminary comment recall that $\tilde{X}$ is a non singular, Vandermonde matrix so that $(\tilde{X}^T\tilde{X})^{-1}$ exists. This is a $(N_a+1) \times (N_a+1)$ matrix that can be computed and stored. Now consider the problem of computing the inverse of $(X^T X)$. To this end note that, from (\ref{Xh}) we have $(X^{(h)})^T X^{(h)} = g_h \tilde{X}^T\tilde{X}$. Then, from (\ref{x2}) we have
\[
X^TX = \left(
\begin{array}{lllll}  
g_1 \tilde{X}^T\tilde{X} & 0 & 0 & ... & 0  \\
0 & g_2 \tilde{X}^T\tilde{X} & 0 & ... & 0  \\
0 & 0 & g_3 \tilde{X}^T\tilde{X} & ... & 0  \\
...  & ...    & ... & ...    &  ...          \\
0 & 0 & 0 & ... & g_{N_g} \tilde{X}^T\tilde{X} 
\end{array} \right)
\]
showing that $X^TX$ is a block diagonal matrix, with $N_g$ blocks and a block dimension of $(N_a+1)$. Such a matrix is easy to invert. Indeed we have
\[
(X^TX)^{-1} = \left(
\begin{array}{lllll}  
\frac{(\tilde{X}^T\tilde{X})^{-1}}{g_1}  & 0 & 0 & ... & 0  \\
0 & \frac{(\tilde{X}^T\tilde{X})^{-1}}{g_2}  & 0 & ... & 0  \\
0 & 0 & \frac{(\tilde{X}^T\tilde{X})^{-1}}{g_3}  & ... & 0  \\
...  & ...    & ... & ...    &  ...          \\
0 & 0 & 0 & ... & \frac{(\tilde{X}^T\tilde{X})^{-1}}{g_{N_g}}  
\end{array} \right).
\]
Then we see that the matrix has a simple structure. It does not need to be really produced and the multiplication by it can be efficently implemented by using only $(\tilde{X}^T \tilde{X})^{-1}$.

\section{Appendix}

{\bf Proof of Lemma 1.} Note that, by construction, $\Sigma_{\hat{X}} \cap \Sigma_P = 0$. Now suppose that the columns of $\hat{W}$ are a dependent set. Then, denoting the columns with $w_i$, we can find coefficients $b_i$ not all zero such that
\[
\sum_i b_i w_i = 0.
\]
Moreover, $w_i = \Pi_P^\perp x_i$ where $x_i$ is the $i$-th column of $\hat{X}$. Since $\Re^n = \Sigma_P \oplus \Sigma_P^\perp$, we can decompose each column as $x_i = x_i^P + x_i^\perp$ where $x_i^P \in \Sigma_P$ and $x_i^\perp \in \Sigma_P^\perp$. The, replacing in the last equation we get
\[
\sum_i b_i \Pi_P^\perp x_i = \sum_i b_i \Pi_P^\perp (x_i^P + x_i^\perp) = \sum_i b_i  x_i^\perp = 0,
\]
where we used the fact that $\Pi_P^\perp x_i^P = 0$ and that $\Pi_P^\perp x_i^\perp = x_i^\perp$. Now consider the vector $v$ that is obtained as a linear combination of the $x_i$ with the coefficients $b_i$, i.e.
\[
v = \sum_i b_i x_i.
\]
Obviously $v \in \Sigma_{\hat{X}}$ and $v \neq 0$ because the coefficients are not all zero. Moreover note that 
\[
v = \sum_i b_i x_i = \sum_i b_i (x_i^P + x_i^\perp) = \sum_i b_i  x_i^P + \sum_i b_i  x_i^\perp = \sum_i b_i  x_i^P.
\]
The last equation shows that $v$ is obtained as the sum of vectors in the signal subspace and therefore $v \in \Sigma_P$. Then $v \in \Sigma_{\hat{X}} \cap \Sigma_P$ which contraddicts the fact that $\Sigma_{\hat{X}} \cap \Sigma_P = 0$. $\diamond$

{\bf Proof of Lemma 2.} Since $\hat{W} = \Pi^\perp_P \hat{X}^{(k)}$ for some $k$, we can prove the lemma if we prove that, for any $k$, the columns of $W$ are in the span of $\Pi^\perp_P \hat{X}^{(k)}$. This is trivially true for $k=0$, because $\hat{X}^{(k)} = X$. Then we can prove the thesis if we show that the columns of $\Pi^\perp_P \hat{X}^{(k)}$ are in the span of the columns of $\Pi^\perp_P \hat{X}^{(k+1)}$. To this end denote the columns of $\Pi^\perp_P \hat{X}^{(k)}$ by $w_i = \Pi^\perp_P x_i$, where the $x_i$ are the columns of $\hat{X}^{(k)}$. Recall that, in the procedure, we look for a vector $v$ which is in $\Sigma_{\hat{X}^{(k)}} \cap \Sigma_P $, so that, using $p_i$ to denote the columns of $P$,
\[
v = \sum_i a_i p_i = \sum_i b_i x_i
\]
and remove from $\hat{X}^{(k)}$ a column $x_j$ such that $b_j \neq 0$. Then we have
\[
v = \sum_i a_i p_i = \sum_{i \neq j} b_i x_i + b_j x_j.
\]
Multiplying the last expression by $\Pi^\perp_P$ we obtain
\[
0 = \sum_{i \neq j} b_i \Pi^\perp_P x_i + b_j \Pi^\perp_P x_j
\]
which yields
\[
w_j = - \frac{1}{b_j} \sum_{i \neq j} b_i w_i
\]
showing that $w_j$ is in the span of the $w_i$ for $i \neq j$. Since the columns of $\Pi^\perp_P \hat{X}^{(k)}$ are the $w_i$ they are in the span of the columns of $\Pi^\perp_P \hat{X}^{(k+1)}$ which are the $w_i$ for $i \neq j$. $\diamond$

{\bf Proof of Lemma 3.} Let us show that the columns of $\hat{X}^{(k)}$ togehter with the columns of $\bar{X}^{(k)}$ are a basis for $\Sigma_X$, for any $k$. Since this is trivially true for $k=0$, we can prove the thesis by showing that, if it is true at step $k$, it is true also at step $k+1$. To this end assume that the columns of $\hat{X}^{(k)}$ togehter with the columns of $\bar{X}^{(k)}$ are a basis for $\Sigma_X$. Note that at the successive step, all the columns will be unchanged except that the $j$-th column of $\hat{X}$, namely $x_j$, is replaced by $v$. But $v$ was selected in the span of $\hat{X}$ i.e. 
\[
v = \sum_i b_i x_i = \sum_{i \neq j} b_i x_i + b_j x_j
\]
so that 
\[
x_j = \frac{1}{b_j}( \sum_i b_i x_i - v ).
\]
Then all the vectors that could be obtained as a linear combination of the vectors $x_i$ for all $i$ (the columns of $\hat{X}$) can also be obtained as a linear combination of the vectors $x_i$ for $i \neq j$ and $v$. This implies that the span of $( \hat{X}^{(k)}, \bar{X}^{(k)})$ is identical to that of $( \hat{X}^{(k+1)}, \bar{X}^{(k+1)})$. In turn this implies that the columns of $( \hat{X}^{(k+1)}, \bar{X}^{(k+1)})$ are a basis for $\Sigma_X$.

Now we can prove the lemma. Based on the last paragraph, we know that the columns of $( \hat{X}, \bar{X})$ are a basis for $\Sigma_X$ so that $\Sigma_X = \Sigma_{\hat{X}} \oplus \Sigma_{\bar{X}}$. Moreover, by construction, we know that $\Sigma_{\hat{X}} \cap \Sigma_P = 0$ and that $\Sigma_{\bar{X}} \subset \Sigma_p$. Then we only have to show that $\Sigma_{\hat{X}} = \Sigma_X \cap \Sigma_P$. Since $\Sigma_{\bar{X}} \subset \Sigma_p$ we only have to prove that all the vectors of $\Sigma_X \cap \Sigma_P$ are also in $\Sigma_{\bar{X}}$. To this end consider a vector $v \in \Sigma_X \cap \Sigma_P$. Since it is in $\Sigma_X$ we can write
\[
v = v_{\hat{X}} + v_{\bar{X}} = \sum_i b_i x_i  + \sum_i c_i h_i
\]
where the $x_i$ are the columns of $\hat{X}$ and the $h_i$ are the columns of $\bar{X}$. Now we note that $v_{\bar{X}} \in \Sigma_X \cap \Sigma_P$ (because the columns of $\bar{X}$ are both in $\Sigma_X$ and in $\Sigma_P$). Then also $(v - v_{\bar{X}}) \in \Sigma_X \cap \Sigma_P$ (because it is a subspace). Moreover 
\[
v - v_{\bar{X}} = \sum_i b_i x_i  
\]
so that $(v - v_{\bar{X}}) \in \Sigma_{\hat{X}}$. From the last equations we have $(v - v_{\bar{X}}) \in \Sigma_{\hat{X}}$ and $(v - v_{\bar{X}}) \in \Sigma_P$ and since $\Sigma_{\hat{X}} \cap \Sigma_P = 0$ we have $(v - v_{\bar{X}}) = 0$. This implies that $v_{\hat{X}} = 0$ and that $v = v_{\bar{X}} \in \Sigma_{\bar{X}}$.  $\diamond$



\begin{thebibliography}{10}



\bibitem{herschel}
G. Pilbratt et al., ''Herschel Space Observatory'', Astronomy and Astrophysics, Vol. 518, No. 7-8, July 2010.

\bibitem{pacs}
A. Poglitsch et al., "The photodetector array camera and Spectrometer (PACS) on the herschel space observatory", Astronomy and Astrophysics, Volume 518, Issue 4, 2010.

\bibitem{spire}
M. J. Griffin et al., "The Herschel-SPIRE instrument and its in-flight performance", Astronomy and Astrophysic, Vol 518, July-August 2010.

\bibitem{tegmark97}
M. Tegmark, ''How to make maps from cosmic microwave background data without losing information'', The Astrophysical Journal, 480, pp. L87-L90, May 1997.



\bibitem{fmri}
N. Bazargani, A.  Nosratinia, K.  Gopinath, R.W. Briggs: "FMRI baseline drift estimation method by MDL principle", Proc. of the IEEE International Symposium on Biomedical Imaging, pp. 472-475, April 2007.

\bibitem{unimap}
http://w3.uniroma1.it/unimap

\bibitem{meyer}
C. D. Meyer: "Matrix analysis and applied linear algebra", SIAM, 2000.

\bibitem{numrec}
W. T. Vetterling, W. H. Press, S. A. Teukolsky, and B. P. Flannery, "Numerical Recipes in C", Cambridge University Press, 1992.

\end{thebibliography}
\end{document}